\documentclass[runningheads]{llncs}
\usepackage[T1]{fontenc}

\usepackage{graphicx}
\usepackage{amssymb}
\usepackage{amsmath}
\usepackage{booktabs}
\usepackage[misc]{ifsym}
\usepackage{multirow}
\usepackage{float}
\usepackage{algpseudocode}

\DeclareMathOperator{\EX}{\mathbb{E}}

\usepackage{epstopdf}
\usepackage{cite}
\usepackage{color,bm}
\usepackage{bbding}
\usepackage{array}
\newcolumntype{C}[1]{>{\centering\arraybackslash}m{#1}}
\begin{document}
\title{Anatomy-Guided Surface Diffusion Model for Alzheimer's Disease Normative Modeling}
\author{Jianwei Zhang \inst{1,2} \and Yonggang Shi  \inst{1,2}}
\institute{Stevens Neuroimaging and Informatics Institute, Keck School of Medicine, University of Southern California (USC), Los Angeles, CA 90033, USA
\and
Ming Hsieh Department of Electrical and Computer Engineering, Viterbi School of Engineering, University of Southern California (USC), Los Angeles, CA 90089, USA
}

\authorrunning{***}
%

%
\maketitle              
\begin{abstract}
Normative modeling has emerged as a pivotal approach for characterizing heterogeneity and individual variance in neurodegenerative diseases,notably Alzheimer's disease(AD). One of the challenges of cortical normative modeling is the anatomical structure mismatch due to folding pattern variability. Traditionally, registration is applied to address this issue and recently many studies have utilized deep generative models to generate anatomically align samples for analyzing disease progression; however, these models are predominantly applied to volume-based data, which often falls short in capturing intricate morphological changes on the brain cortex. As an alternative,surface-based analysis has been proven to be more sensitive in disease modeling such as AD, yet, like volume-based data, it also suffers from the mismatch problem.To address these limitations, we proposed a novel generative normative modeling framework by transferring the conditional diffusion generative model to the spherical non-Euclidean domain. Additionally, this approach generates normal feature map distributions by explicitly conditioning on individual anatomical segmentation to ensure better geometrical alignment which helps to reduce anatomical variance between subjects in analysis. We find that our model can generate samples that are better anatomically aligned than registered reference data and through ablation study and normative assessment experiments, the samples are able to better measure individual differences from the normal distribution and increase sensitivity in differentiating control normal (CN), mild cognitive impairment (MCI), and Alzheimer's disease (AD) patients.


\end{abstract}
\section{Introduction}
Normative modeling has been proven to be an effective scheme for modeling neurodegenernative diseases such as Alzheimer's disease \cite{10.7554/eLife.85082}. The core idea of normative modeling is defining normal distribution such that each subject can be measured against it to characterize deviation from norm. One of the major challenges of such tasks is the individual anatomical variability. Specifically, the cortical folding patterns exhibit considerable heterogeneity across individuals, thereby complicating the establishment of a meaningful comparative baseline.  Conventionally, statistical analysis techniques are applied on the anatomically registered images to attenuate effect of individual variability. However, due to shape differences, registered images still have significant gyral/sulcal mismatch \cite{10.1007/978-3-031-43904-9_6} and statistical methods are usually limited in their abilities to capture complex nonlinear relationships. 
\par As an alternative, deep generative models have recently been introduced to address these limitations. The idea is to train a generative model that encodes how normal distribution behaves.
Variational Autoencoder (VAE) \cite{ICAM,RAVI2022102257},flow-based model \cite{9008303},Generative Adversarial Network (GAN) \cite{BAI2022353} are employed to model the normal distribution on the brain MRI volume space and utilize deviation of original data from generated normal samples as disease atrophy map for analysis. Although, these previous research achieved good results on the volume data, few attempts have been made to adapt these method for cortical surface-based data, which has been proven to more prominent at capturing detailed anatomical changes. \cite{HUTTON2009371,LERCH2005163}.

\par Traditional surface-based analysis are built upon registered brain surfaces with average surface or preset atlas, which also suffers from cortical structure mismatch. To account for this problem, previous researches have attempted personalized analysis where, instead of using entire dataset, only a subset of similar anatomical structures are used for analysis\cite{10.1007/978-3-031-43904-9_6,10.1007/978-3-030-87234-2_67}. However, these approaches suffers from limited data acquisition and computational complexity as high cortical variability might not represented by the existing datasets. Therefore, generative model is a promising alternative approach to generated personalised reference set instead of trying to match against real data. 
\par Recently, diffusion model has emerged as an effective model for stable and effective image generations. To leverage this advancement of generative model, in this paper, we adapt the Denoising Diffusion Probabilistic Models(DDPM) framework \cite{ho2020denoising} from euclidean image domain to non-euclidean surface domain and propose a conditional surface diffusion model that utilize gyral sulcal segmentation mask to generate cortical surface features that are anatomically aligned. The proposed model is applied to ADNI dataset\cite{MUELLER2005869} to conduct ablation study and normative modeling on control normal(CN),mild cognitive impairment (MCI) and Alzheimer's disease(AD) subjects. The results show that our model is able to generate faithful and anatomical aligned feature map and increase the sensitivity of surface based disease analysis. 

\section{Method}
\subsection{Denoising Diffusion Probabilistic Models(DDPM)}
DDPM \cite{ho2020denoising,rombach2022highresolution} is a kind of iterative generative model for modeling data distribution from samples. Given a series of observed samples {$x_{i}$}, which is drawn from the data distribution p(x), model learns to generate new samples from p(x) through a forward and backward diffusion process. The diffusion process of DDPM is governed by a Markov chain as in equation \ref{ddpm_forward},\ref{ddpm_backward}, describing forward and backward process respectively.
\begin{equation}
    q(x_{t}|x_{t-1}) = \mathcal{N}(x_{t}; \sqrt{1-\beta_{t}}x_{t-1},\beta_{t}\mathcal{I}) \quad q(x_{1:t}|x_{0}) = \prod_{t=1}^{T} q(x_{t}|x_{t-1})
    \label{ddpm_forward}
\end{equation}
\begin{equation}
    p_{\theta}(x_{0:T}) = p(x_{T}) \prod_{t=1}^{T} p_{\theta}(x_{t}|x_{t-1}), 
    p_{\theta}(x_{t}|x_{t-1}) = \mathcal{N}(x_{t-1}; \mu_{\theta}(x_t,t),\Sigma{\theta}(x_t,t))
    \label{ddpm_backward}
\end{equation}

$x_{0}$ is the original data and $x_{t}$ is the noisy data after adding t steps of noise. The T denotes the total number of steps. $\beta_{t}$ is from a predefined set of variance schedule $\{\beta_{t} \in (0,1)\}|_{1}^{T}$. The information within the data is progressively destroyed by adding independent Gaussian noise for a certain number of steps in the forward process. The backward process is then formulated as a sampling process by implementing a neural network to estimate $\mu_{\theta}(x_t,t),\Sigma{\theta}(x_t,t)$ iteratively and denoise the noisy data  $x_{T}$ to achieve new sample generation. \\
\begin{equation}
    L = \EX_{t  \thicksim [1,T],x0,\epsilon_{t} }\Bigr[ ||\epsilon_{t} - \epsilon_{\theta}( \sqrt{\overline{\alpha}_{t}}x_{0} + \sqrt{1-\overline{\alpha}_{t}}\epsilon_{t},t)|| \Bigr]
    \label{eq:loss}
\end{equation}
\par The model is trained by optimizing a simplified Evidence Lower Bound loss\cite{ho2020denoising} as shown in equation \ref{eq:loss}. $\epsilon_{t}$ is the noise at time t and $\epsilon_{\theta}$ is the neural network for estimating noise at each time step. $\overline{\alpha}_{t}$ is $\prod_{i=1}^{T}{(1-\beta_{i})}$. We employed the cosine beta schedule as in \cite{nichol2021improved} as the variance schedule and the velocity sample update scheme in \cite{salimans2022progressive}, which we empirically find it to be more stable than the original scheme. The training procedure is same as \cite{ho2020denoising}. After training, the DDPM model is able to generate new samples from p(x) from either random noise or noisy input data. 

\begin{figure}[tb]
    \centering    
    \includegraphics[width=\textwidth]{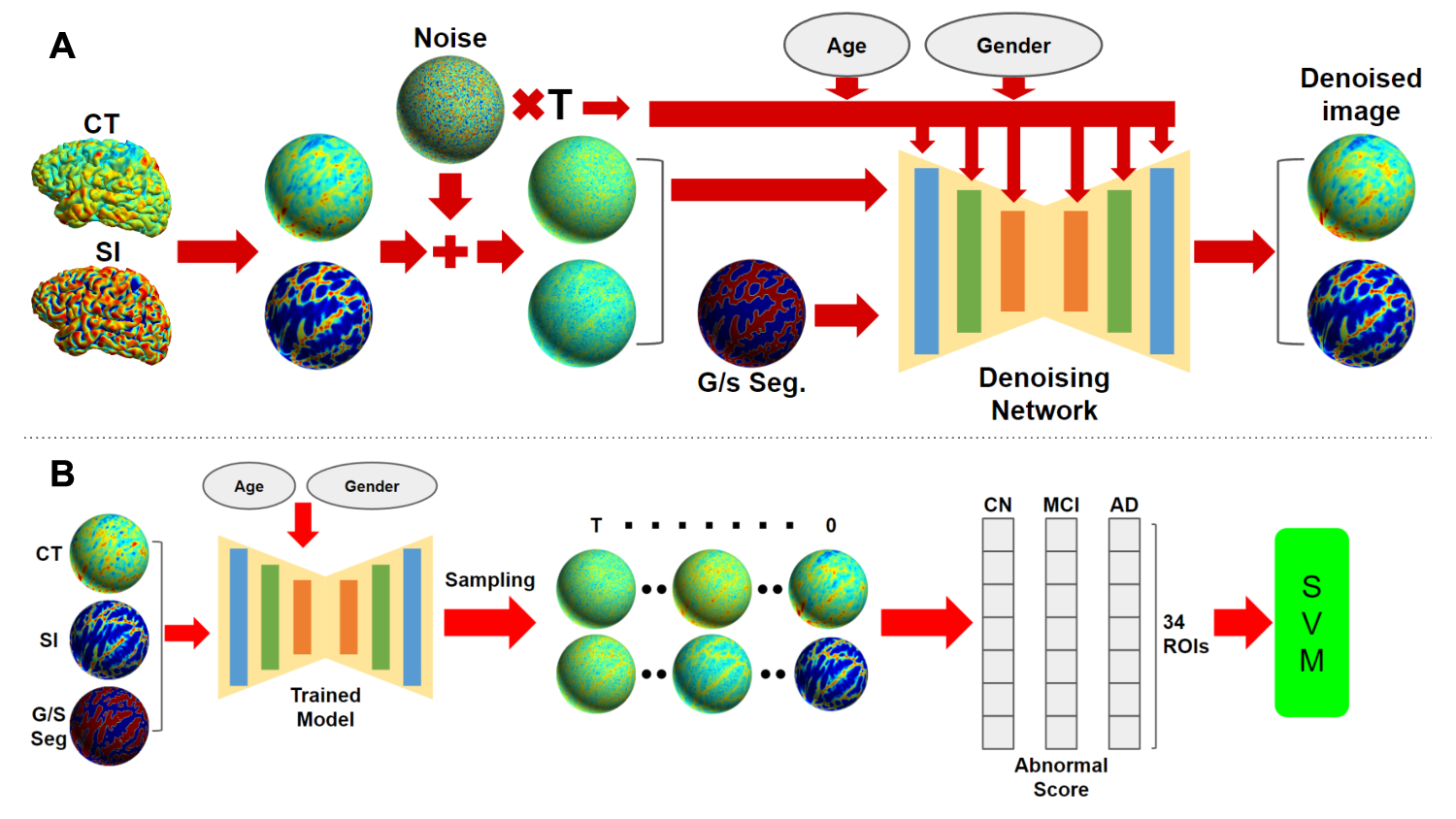}

    \caption{(A) The training procedure of the conditional DDPM model, where CT denotes cortical thickness, SI denotes shape index and G/S seg. denotes Gyral/Sulcal segmentation (B) The pipeline of sampling process for each test subject to genrate abnormal score for analysis (Note: all images are real data and real generated samples)}
    \label{fig:pipeline}
\end{figure}

\subsection{Denoising Network in Spherical Domain} 
\begin{figure}[tb]
    \centering    
    \includegraphics[width=\textwidth]{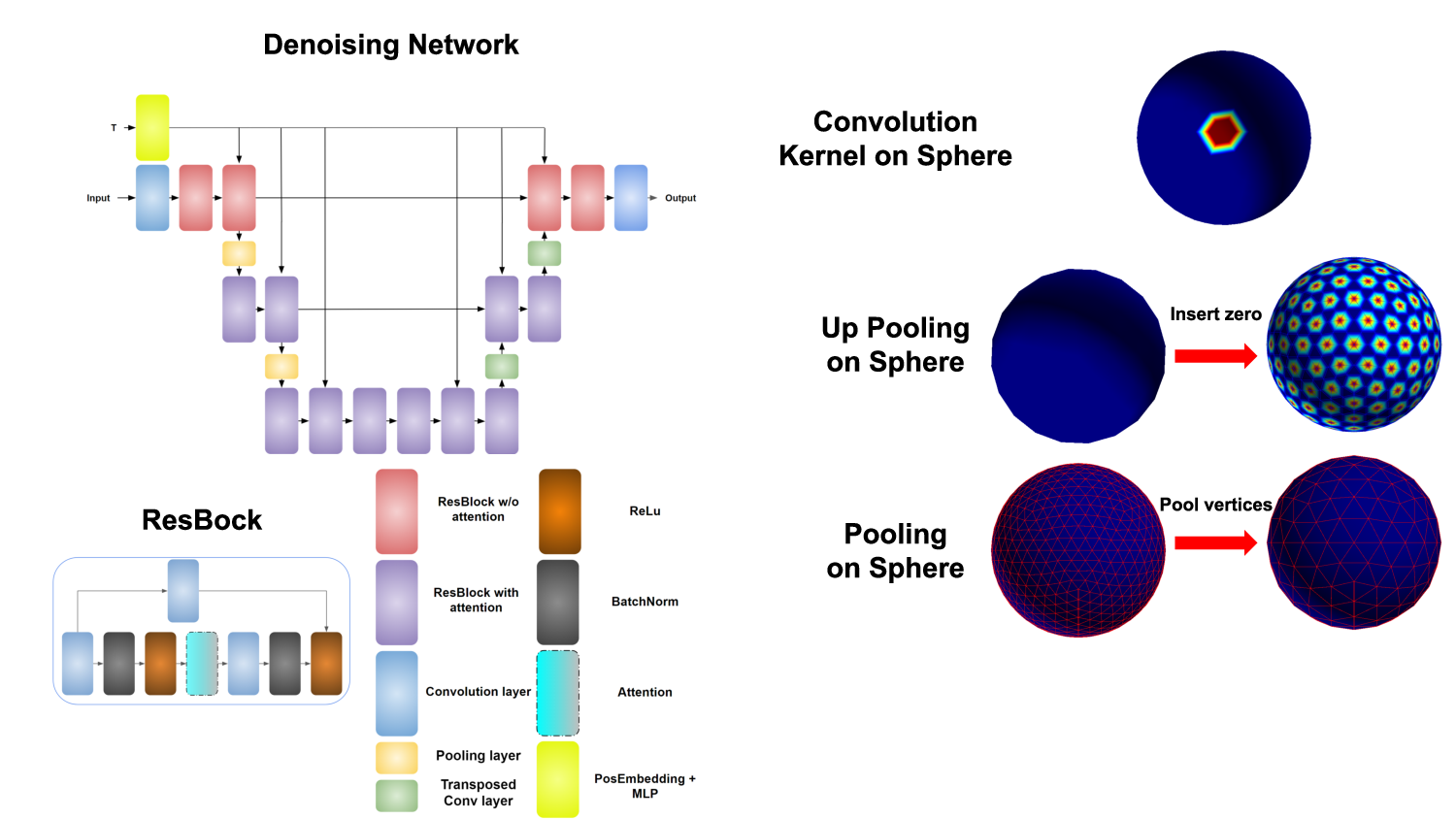}

    \caption{The figure shows the structure of the denoising network which has a UNet structure with ResBlock built out of spherical convolution, pooling, up-pooling and attention layer.}
    \label{fig:network}
\end{figure}

To align data in a common space, the feature maps are resampled to a standard icosahedron. Unlike in euclidean domain, there is no native definition of direction in the spherical domain. In order to mirror the operation of convolution in 2d image domain, we adapt the convolution method from \cite{Spherical_unet}, which defines convolution on the icosphere by the 1 ring neighborhood.
The original Unet structure in \cite{Spherical_unet} utilize a neighborhood averaging strategy for pooling and up-pooling, however, empirically, we find that such operations introduce grid-like artifacts into the generated samples. To reduce artifacts and mimic operations in euclidean domain, we employ a different pooling and up pooling methods. Utilizing the natural structure of the icosahedron, the pooling for ith order can be defined as only keeping vertices in the (i-1)th order icosahedron and up pooling is the zero padding for vertices added from ith to (i\textbf{+}1)th order icosahedron. Figure \ref{fig:network} shows the overall structure of the network and illustrations of the operations. 

\subsection{Anatomical and Demographic Conditioning}
The original DDPM is for modeling unconditional feature distribution. In order to better model the normal distribution of cortical features based on anatomical structure, we modified the model to take additional conditioning. In our configuration, there are 2 types of conditions used in training conditional DDPM: demographic and anatomical condition. The demographic conditions include gender and biological age of each subject. Both values are first passed into the network through vector embedding layers. The vector embedding are then added to the time embedding \cite{ho2020denoising} and passed to each ResBlock in the network. For anatomical condition, which is in the form of a binary segmentation mask\cite{4389763}, is concatenated to the input feature map to insert into the network. 

\subsection{Sampling for Normative Modeling}
The core idea for our normative modeling is to use sampled feature map to measure deviation score as opposed to directly using real data because of the anatomical mismatch. Through procedures described in previous section, the model will generate N samples per test subject conditioned on original image with 500 steps of added noise, which is determined empirically, individual anatomical segmentation, gender and age information. For each Region of Interest(ROI) defined by FreeSurfer output file \textit{aparc.annot}, an abnormal score is computed as in equation \ref{eq:zscore}. 

\begin{equation}
    Z_{i} = \frac{x_{i}-mean([x_{(i,1)}...x_{(i,N)}])}{std_{j}([x_{(i,1)}...x_{(i,N)}])}
    \label{eq:zscore}
\end{equation}
For one test subject, $Z_{i}$ is the abnormal score for ith ROI. N is the number of samples. $x_{i}$ is the mean feature value of the test subject in ith ROI. $x_{(i,j)}$ denotes the mean feature value for ith ROI of jth sample. The abnormal scores measures how much does a test subject deviate from normal in each ROI. Additionally, the abnormal scores of 34 ROIs are formed into 1 by 34 vector and passed into a standard SVM for 10-fold cross validation of CN vs MCI, CN vs LMCI and CN vs AD classification.

\section{Experiments and Results}
\subsection{Preprocessing}
We use the Alzheimer’s Disease Neuroimaging Initiative(ADNI) dataset\cite{MUELLER2005869} to conduct experiments. 646 subjects are selected, including 482 control normal (CN), 82 mild cognitive impairment (MCI), and 82 Alzheimer’s disease (AD) patients. 400 CN subjects are used as training set and all other subjects are used as testing set. All the T1 MRI volume images are processed through FreeSurfer 6.0 \cite{DALE1999179} to extract the cortical surfaces and cortical thickness map. The shape index map and gyral/sulcal segmentation mask are obtained following \cite{4389763}. All feature maps and segmentation mask are resampled to a standard 6th order icosahedron of 40962 vertices using \textit{mris\_surf2surf}command. To normalize the data, cortical thickness map is scaled to between -1 and 1 using min of 0 mm and max of 5 mm. Gender label is represented by scalar 0 and 1. Age label is scaled to [0,1] range by dividing by 100. For computational cost, the experiments are only conducted on the left hemisphere. 

\subsection{Implementation Details}
All feature maps and segmentation masks are 1 $\times$ 40962 vectors. The input to the network is the concatenation of cortical thickness map and shape index map with the gyral/sulcal segmentation mask in 1 hot encoding, resulting in a matrix of size B $\times$ 4 $\times$ 40962, where B is batch size. The max timesteps of DDPM is set at 1000. The model is trained with ADAM as optimizer, cosine annealing as scheduler and a starting learning rate of 1e-5 for a total of 1000 epochs, about 24 hours. The network is implemented using Pytorch and trained on a NVIDIA A5000 GPU.

\subsection{Ablation Study for Conditional DDPM}
We performed ablation study on the conditioning introduced to DDPM to demonstrate improvement in generated samples. The experiment is conducted on the CN subjects' cortical surface. Two models are trained on the 400 CN subject: unconditional DDPM and DDPM with gyral/sulcal segmentation. All test data are first blurred with a T value of 500, then denoised for reconstruction. The test is conducted on the 82 test CN subjects. The mean SSIM and mean MSE between reconstructed samples and real data are shown in table \ref{tab:ab}. As reference, we also computed the mean SSIM between test subject and 1 randomly selected registered CN subject to show that the model can generated better aligned samples comparing to registered real data. Through the ablation study, the conditioning can indeed improve the reconstructed sample quality and produce better aligned feature maps. 

\begin{table}[tb]
\makebox[\textwidth]{
\begin{tabular}{|c|c|c|}
\hline
Model type & SI SSIM$\uparrow$  & SI MSE $\downarrow$ \\
\hline
Template data & -0.0102 $\pm$ 0.0014 & N/A\\
\hline 
Unconditional DDPM  & 0.4911 $\pm$ 0.0165 & 0.1461 $\pm$ 0.0064\\
\hline
DDPM + segmentation mask & \textbf{0.6167 $\pm$ 0.0127} & \textbf{0.1011 $\pm$ 0.0046}\\
\hline
Model type & CT SSIM$\uparrow$ & CT MSE(mm) $\downarrow$\\
\hline
Template data & 0.2311 $\pm$ 0.0136 & N/A\\
\hline 
Unconditional DDPM  & 0.3978 $\pm$ 0.0168 & 0.4329 $\pm$ 0.0174\\
\hline
DDPM + segmentation mask & \textbf{0.4903 $\pm$ 0.0224} & \textbf{0.3894 $\pm$ 0.0253}\\
\hline
\end{tabular}
}
\caption{This table shows that results of ablation study}
\label{tab:ab}
\end{table}

\subsection{Normative Assessment on ADNI dataset}
To evaluate model's performance on reducing heterogeneity introduced by the anatomical mismatch, we compare our model to spherically registered real data using FreeSurfer \cite{DALE1999179}. The conditional DDPM model is trained on the 400 template CN subjects and after training, for each CN,MCI and AD subjects in the test set, 10 samples are generated as DDPM reference set. For fair comparison, 10 subjects whose age is closest to the test subject are picked as the template reference set.. Both sets are used to compute abnormal score per subject per ROI as described in the method section. 

\begin{figure}
    \centering    
    \includegraphics[width=\textwidth]{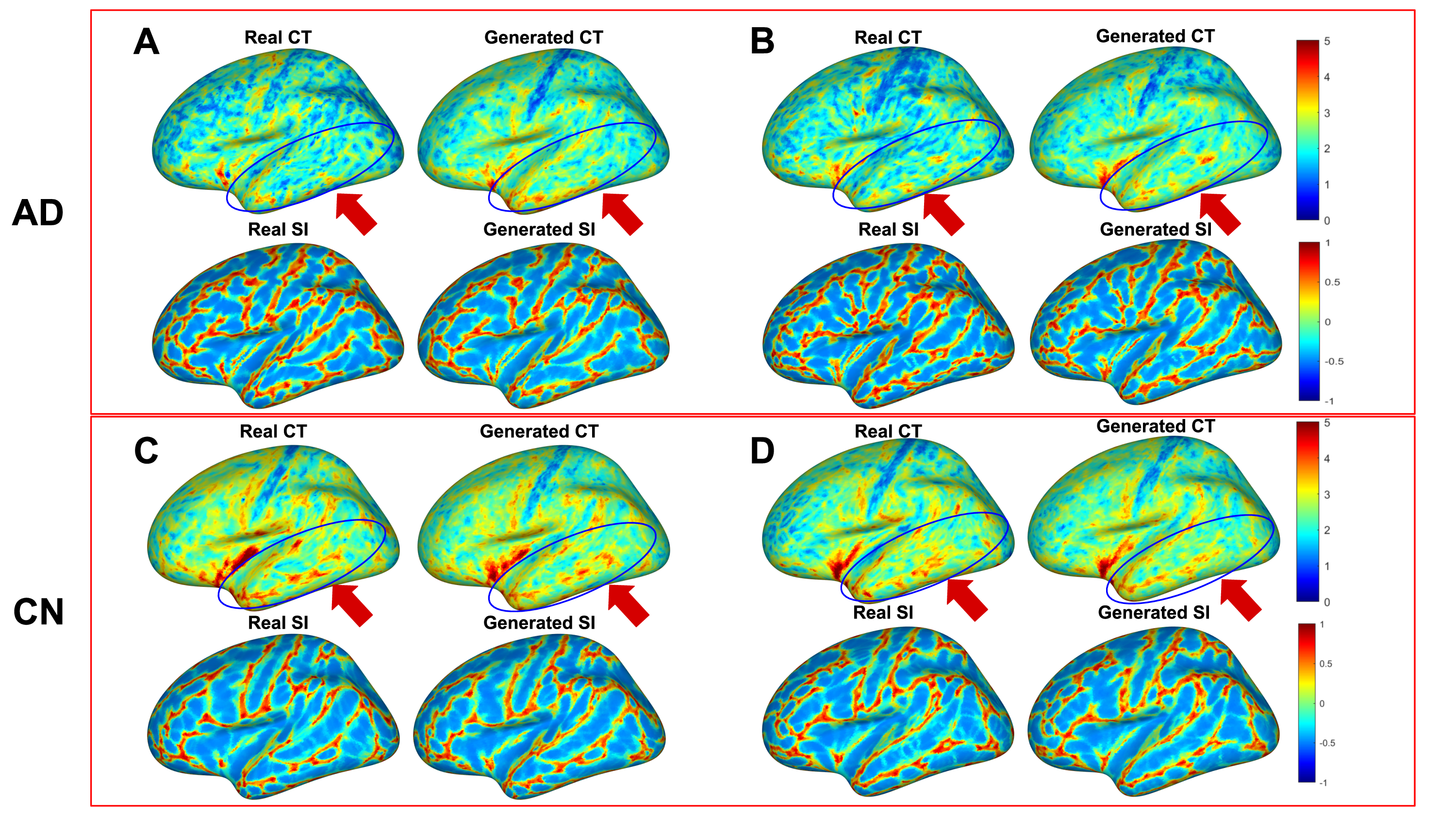}

    \caption{This figure shows the comparison between real and generated feature maps. CT denotes cortical thickness. SI denotes shape index. Figure A, B are AD subjects and C, D are CN subjects. The blue circles highlight the temporal region which highly correlates with AD. }
    \label{fig:real_gen_compare}
\end{figure}

\begin{figure}[tb]
    \centering    
    \includegraphics[width=\textwidth]{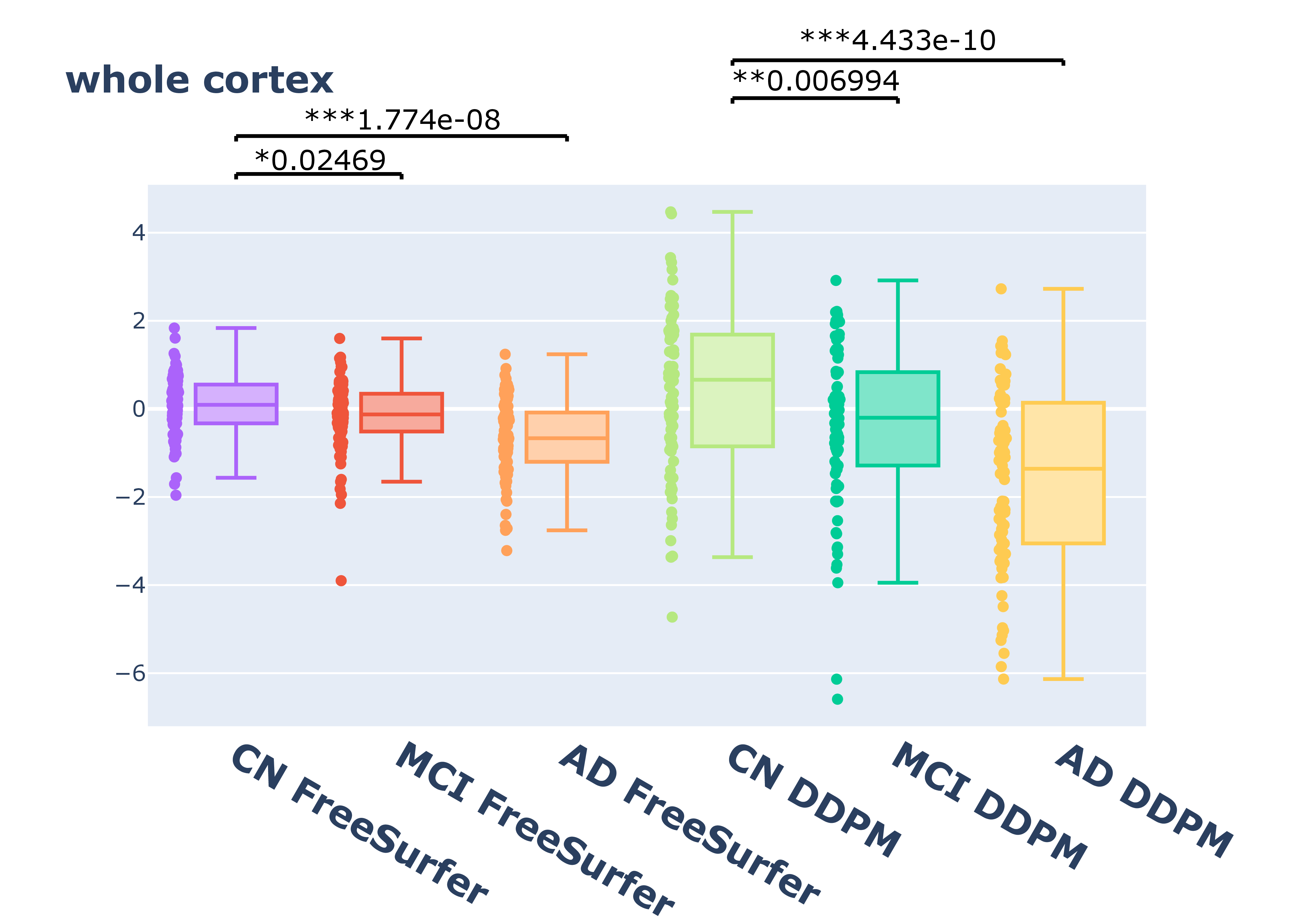}

    \caption{The figure shows the distribution comparison between abnormal score per subject of the whole cortex for template and DDPM reference sets.}
    \label{fig:norm_test}
\end{figure}

A qualitative comparison of real and generated data is shown in fig \ref{fig:real_gen_compare}. The figure shows that our model trained on CN subject is able to estimate the normal distribution of an AD subject based on the individual anatomy,especially in the temporal region which correlates with AD. Figure \ref{fig:norm_test} shows the box plot of the mean abnormal score across ROIs of the whole cortex. The p-values between CN vs MCI and CN vs AD are computed for both reference sets and shown in the figure. Based on the p-values, our model produces abnormal scores with better sensitivity for CN vs MCI and CN vs AD. \\

\begin{table}[tb]
\makebox[\textwidth]{
\begin{tabular}{|c|c|c|c|c|c|c|c|}
\hline

\multicolumn{4}{|c|}{CN vs AD} & \multicolumn{4}{|c|}{CN vs MCI}\\
\hline
Score Type & Accuracy  & Precision & Recall & Score Type & Accuracy  & Precision & Recall \\
\hline
Template & 0.7529 & 0.7897 & 0.7458 & Template & 0.6058 & 0.6211 & 0.5875\\
\hline
DDPM  & \textbf{0.8117}  & \textbf{0.8310} & \textbf{0.8041} & DDPM  & \textbf{0.7117}  & \textbf{0.6819} & \textbf{0.7986}\\
\hline
\end{tabular}
}
\caption{The table shows the accuracy, precision and recall of classification on CN vs MCI and CN vs AD}
\label{tab:class}
\end{table} 
\par In addition to distribution level comparison, we also conducted classification experiment for CN vs MCI and CN vs AD. The abnormal score is computed for 34 ROIs for all subjects, which is formatted as a length 34 vector. This vector is then used as feature for classification in a standard SVM classifier. To validate the results, we have performed 10-fold cross validation with train test ratio of 1:9 and the accuracy,precision and recall are shown in table \ref{tab:class}. In both CN vs AD and CN vs MCI, our abnormal score performs better than template's.





\section{Conclusion}
In this paper, we proposed a framework for DDPM model on the spherical domain, conditioned on the anatomical segmentation to generate geometrically alignment feature maps. The ablation study and normative tests have shown that our model can generate reliable feature maps on the cortical surface and perform better than registered reference set in normative modeling of AD. 

\clearpage
\newpage
\bibliographystyle{unsrt}
\bibliography{references}
\end{document}